# Robust Downlink Throughput Maximization in MIMO Cognitive Network with more Realistic Conditions: Imperfect Channel Information & Presence of Primary Transmitter.

**Adnan Gavili Gilan[*], Masoumeh Nasiri-Kenari[**]**

***adnan.gavili@gmail.com ** mnasiri@sharif.edu**

Abstract: *Designing an efficient scheme in physical layer enables cognitive radio (CR) users to efficiently utilize resources dedicated to primary users (PUs). In this paper in order to maximize the SU's throughput, the SU's transceivers beamforming is designed through new model considering the presence of the PU's transmitter. Since presence of primary transmitter basically degrades CR's system performance; proposed beamforming design considers intra-system interference between PUs and SUs. An optimization problem based on maximizing CR network throughput subject to controlling interference power from SU transmitter to PU receiver has been formulated. Due to limited cooperation between PU and SU network, channel state information (CSI) between two networks are assumed to be partially available, subsequently conventional CSI uncertainty model known as norm bounded error model has been employed. The proposed optimization problem, which is basically difficult to solve, has been converted to a semi definite program which can be efficiently solved by optimization toolbox software e.g., CVX-Mathlab. Furthermore, alternative time efficient and close form solutions are derived. The superiority of the proposed approach in comparison with the previous works has been confirmed through the simulation results.*

## I. Introduction

Traditionally, fix spectrum resource allocation has been deployed for many applications throughout the past years [1]. Experimental results demonstrate that previous allocated spectrums have not been utilized efficiently [2], [3], [4], [5]. In order to mitigate fixed spectrum allocation problems, the concept of cognitive radio is attacked in [6], which enables some lower priority users, called secondary users (SUs), to utilize the spectrum resources dedicated to high priority primary users (PUs). The SU's can utilize spectrums in two ways: 1) If the PU's spectrum is vacant in a great portion of the time, the SU can transmit its data in this media, but if the PU wants to use its spectrum, PU must run a handoff procedure to leave the frequency band and try to capture a channel to continue its unfinished transmission. This case is so called opportunistic spectrum access [1]. 2) If the spectrum is usually busy by the PU, another scheme is that to utilize special transmission opportunities made by beamforming concept. In this case, the SU simultaneously transmits its data in the direction of the SU's receiver and guaranties the interference to the PU receiver to be lower than a predefined threshold. This case is called concurrent spectrum access [7]. The challenge of how to mitigate spectrum inefficiency has led many pioneers to provide new methods to overcome spectrum scarcity problem. The major works on throughput maximization which has been conducted before were to propose methods on how to sense and find spectrum holes, which comes to some specific researches mainly on spectrum sensing and spectrum handover [1]. These contributions are mainly valuable for usually vacant spectrums or time holes in white spaces because the chance for the CR to find temporal spectrum holes would be very high. Although finding transmission opportunities in areas with dense distribution of PUs is mainly impossible; MIMO beamforming scheme leads to a better solution since it provides a method to send data in a specific direction in special opportunities. Regarding this idea, the SU would be able to use non-vacant spectrums by simultaneously transmitting its data in one direction and controlling interference to the PU's receiver direction. Using MIMO beamforming essentially enhances SU performance in highly dense areas. In the beamforming concept, some common parameters become more important, e.g., signal to interference plus noise ratio (SINR), cognitive signal to interference ratio (CSIR) and interference power (IP) [8], [9], [10], [11], [12].

Recently, using multiple antennas has gained a lot of interest in cognitive radio networks (CRNS). In [8], a primary single antenna network and secondary MISO transmitter-receiver is considered. Assuming a perfect knowledge of all CSI information, CSIR maximization through deriving optimum beamforming vector at SU transmitter is investigated. While maximizing CSIR, the total interference to PUs was kept lower than a predefined threshold (interference limit) but individual interference to PUs is not taken into account. Further, only the effect of the SU transmitter on the PU receiver is investigated and the effect of PU transmitter on SU receiver is not considered. Considering statistically CSI and norm bounded error (NBE) model, the authors in [7] try to maximize the expected SNR at the SU receiver in which transmitting and receiving beamforming is derived through a second order cone program (SOCP). In this paper, partially-feedback CSI is handled and a SOCP problem is proposed, which can be effectively solved by optimization software packages. Designing a robust transmitter beamforming for MISO cognitive pair has been investigated in [9]. Taking the impact of limited CSI into account, the authors derived a close form solution in which satisfy all channel realization in considered CSI model .In [10], [11], a beamforming transmitter scheme in the broadcast channel (BC) has been proposed, and the authors converted the problem in to a semi definite program (SDP). In [12], [13], [14,] a robust beamforming design for MIMO BC channel is proposed in which the authors aim to minimize the mean square error (MSE) through appropriately design the transceivers beamforming matrices.

In this paper, concurrent MIMO SU network has been placed at the point of major concentration where SU transceiver beamforming vectors has been designed to maximize the SU network throughput in the downlink case. Having all CSIs between SU transmitter and SU receiver completely and partial CSIs between PUs and SUs, in this paper, the problem of robust beamforming design is formulated. Further, an optimization problem is formulated in order to jointly maximize SU network throughput and control the interference to PU receiver. Since PU transmitter may cause destructive interference on SU receiver, receiver beamforming at SU is considered in order to mitigate (full CSI available) or control (partially CSI available) interference from PU. To solve the optimization problem, it is converted into a SDP form and then solved by convex optimization toolboxes, e.g., CVX [18]. In order to mitigate the huge computational burden imposed by exhaustive search, a close form solution has been derived for single user case. For multiple SUs, two different solutions are proposed and the pros and cons of each aspect are discussed in detail. First, OFDM-beamforming concept is considered to share a common resource among SUs. The whole spectrum dedicated to a PU is decomposed into non-overlapping sub bands, and each SU could transmit on one of them with its optimal beamforming design. In such circumstances, optimum close form solution and a descriptive algorithm are provided to solve the problem. Second, the concept of underlay transmission, e.g., UWB, has taken into account in order to increase the throughput of the SU network, where, all SUs could transmit concurrently at the PU's spectrum. The solution of the optimization problem in this case may not be easily derived in a close form; therefore a SDP solution has been proposed. While second method enhances the network throughput of SU in compare with the first one, the time needed for solving the optimization problem in the former method is weigh more than that of the latter one.

The rest of the paper is organized as follows. In Section II, a novel system model for a pair of SU transmitter-receiver has been introduced, and an optimization problem in the presence of primary transmitter interference to the secondary receiver has been formulated. For single user system, considering imperfect CSI, optimal transmitter and receiver beamforming vectors are jointly designed in Section III. Furthermore, a SDP relaxation and close form solution are presented to efficiently solve the derived optimization problem. This model is generalized to multi user secondary in Section IV. Section V presents the simulation results, and further discussions are outlined as well. Section VI includes conclusion.

*Notations*

In this paper matrices and vectors are respectively denoted by upper and lower case character. £ is a set of complex numbers. $(.)^T$, $(.)^H$ are transpose and complex conjugate transpose of a vector or matrix. $tr[.]$ represents the trace of a matrix, and $\|.\|$ is a norm operator. If a matrix is positive semi definite, it will be shown as $A \geq 0$.

## II. System model an problem formulation

It is assumed that the SUs and PUs transceivers subsequently have multiple and single antennas. General framework of this scheme is shown in Fig.1.

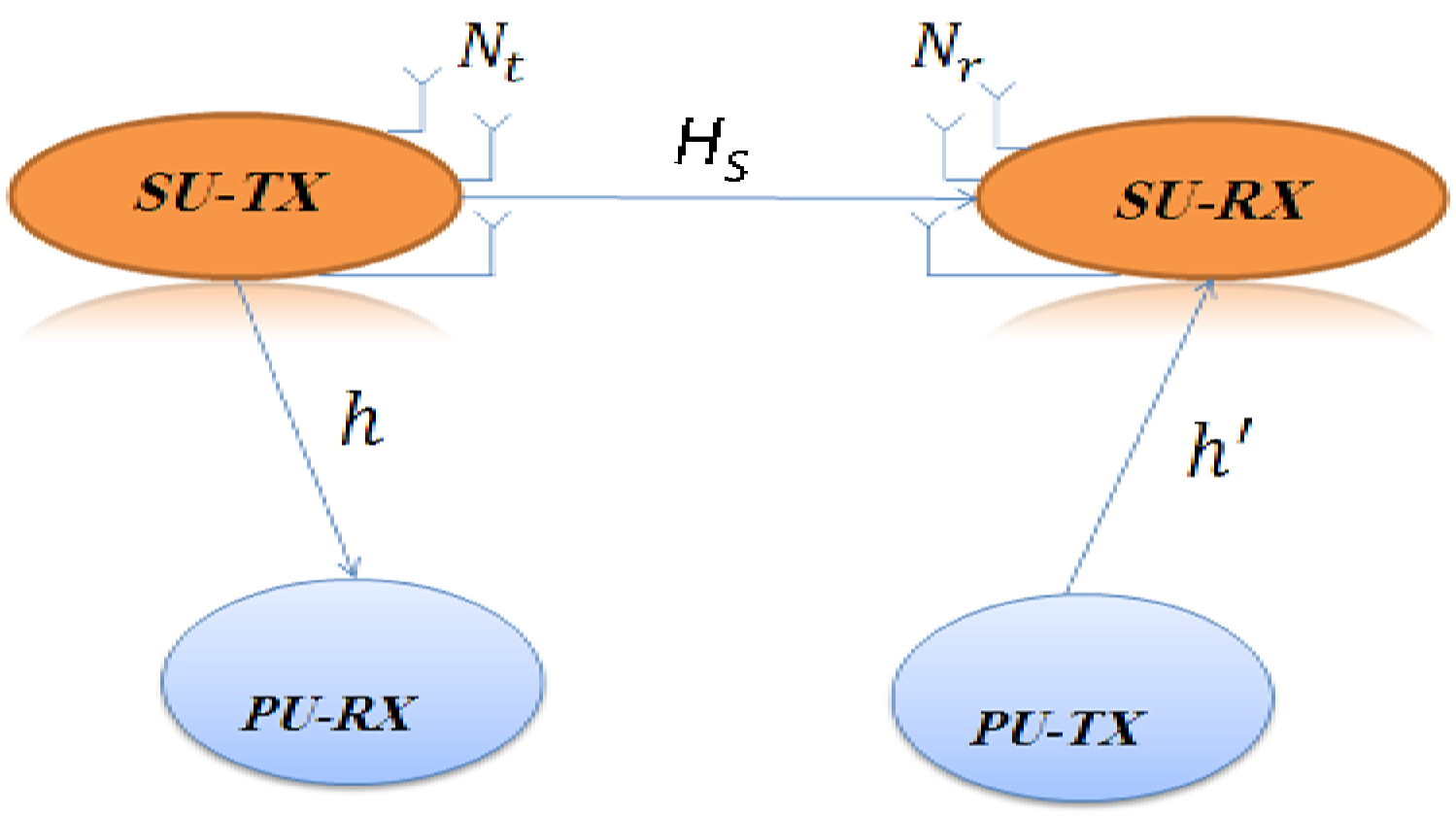

Fig.1. Proposed model

It is defined that $H_s \subset \pounds^{N_r \times N_t}$ is channel matrix between SU transmitter-receiver pair which is assumed to be completely known in advance and $h^{'} \subset \pounds^{N_r \times 1}$ is channel vector between PU transmitter and SU receiver which is partially known as mentioned before [15]. While the channel matrix between SU's transmitter and receiver, $H_s$, is precisely available, as it is demonstrated in the Fig.1., in order to properly model the limited cooperation between the primary network and the CRN, the CSI between SU transmitter and PU receiver as well as PU transmitter and SU receiver is assumed partially available. Generally, for a SU/PU pair in each transmission, transmits symbols $s_i / s_i^{'}$ are assumed to be zero mean independent and identically distributed (i.i.d.). Also the transmission symbols in consecutive order are considered to be independent.

$$\begin{aligned} &E\left(s_i s_j^{*}\right) = P_2\, d\left(i - j\right) \\ &E\left(s_i^{'} s_j^{'*}\right) = P_1\, d\left(i - j\right) \qquad (1) \\ &E\left(s_i s_i^{'}\right) = 0 \end{aligned}$$

where $P_1$ and $P_2$ are the SU and PU transmit power limit. Since symbols need a medium to be transmitted, transmitting antennas were weighted with different beamforming vector in each transmission attempt in order to send the data in a specific direction. The transmitted signal after applying beamformer is [17]:

$$x_i = s_i \, w_i \in \mathbb{C}^{N_t \times 1} \tag{2}$$

where the vector $w_1$ represents transmitting beamformer. At the SU receiver the received signal is:

$$y_i = s_i \times H_s \times w_1 + s_i^{'} \times h^{'} + n_i \tag{3}$$

where $n_i \subset \mathbb{C}^{N_r \times 1}$ is the noise vector which has been considered to be element wise independent and also independent from transmitting symbols.

$$\begin{aligned} &E\left(n_i\right) = 0 \\ &E\left(n_i n_j^{H}\right) = \sigma^{2}_{Noise} I_{N_r} \end{aligned} \tag{4}$$

Applying receiver beamformer, $W_2$ , to the received signal, we have.

$$z_i = w_2^{H} y_i = s_i \times w_2^{H} \times H_s \times w_1 + s_i^{'} \times w_2^{H} \times h^{'} + w_2^{H} \times n_i \tag{5}$$

The above equation comprises three different and statistically independent parts. 1) SU signal, 2) interference caused by PU and 3) the noise component. Interfering signal to primary receiver can be described as follow.

$$z_i^{'} = s_i \times w_1^{H} \times h \tag{6}$$

As [9] and [12], the NBE model is exploited in order to consider the partially available CSI. The CSI between PU/SU transmitter and SU/PU receiver is as follows.

$$
\begin{aligned}
&(h-h_0)R^{-1}(h-h_0)^H \le e \\
&(h'-h_0')R'^{-1}(h'-h_0')^H \le e
\end{aligned}
\tag{7}
$$

where correlation matrix $R/\,R'$ is the channel error correlation for $h/\,h'$ and $e$ is the amount of error in CSI model [9]. While it is assumed that the instantaneous feedback of channel vector $h/\,h'$ is not available, the mean CSI are available [9].

$$
\begin{aligned}
&R = E\left\{(h-h_0)\times(h-h_0)^H\right\} = s^2 \times I \\
&R' = E\left\{(h-h_0)'\times(h-h_0)'^H\right\} = s^2 \times I
\end{aligned}
\tag{8}
$$

Those sets of formula in equation (7) are easily converted to the following.

$$
\begin{aligned}
&\left|h-h_0\right| \le s\sqrt{e} \\
&\left|h'-h_0'\right| \le s\sqrt{e}
\end{aligned}
\tag{9}
$$

The following equations are easily verifiable.

$$
\underset{s_i,\, s_i',\, n_i}{E}\left(z_i^2\right) = E\left(s_i^2\right)\times\left|w_2^H \times H_s \times w_1\right|^2 + E\left(s_i'\right)\times\left|w_2^H \times h'\right|^2 + w_2^H \times E\left(n_i n_i^H\right)\times w_2 \tag{10}
$$

$$
E\left(s_i^2\right) = P_2 \;\&\; E\left(s_i'\right) = P_1 \;\&\; E\left(n_i n_i^H\right) = s^2_{Noise} I_{N_r} \tag{11}
$$

$$
P_{Int} = \underset{s_i}{E}\left(z_i'^2\right) = E\left(s_i^2\right)\times\left|w_1^H \times h\right|^2 \tag{12}
$$

$$
P_S = \underset{s_i}{E}\left(x_i^2\right) = E\left(s_i^2\right)\times w_i^H \times w_i \tag{13}
$$

where $P_{Int}$ and $P_S$ are respectively the interference power to PU receiver and SU transmitted power. Therefore, the SINR can be easily computed.

$$
SINR = \frac{P_2 \times \left|w_2^H \times H_s \times w_1\right|^2}{P_1 \times \left|w_2^H \times h'\right|^2 + s^2_{Noise} \times w_2^H \times w_2} \tag{14}
$$

SU's rate based on Fig.1 is derived as follows (See appendix A-1).

$$R = \log_2\left(1 + \frac{w^H{}_2\left(pH_S w_1 w^H{}_1 H_S{}^H\right)w_2}{w^H{}_2\left(p'h'h'^H + s_{Noise}{}^2 I\right)w_2}\right) = \log_2\left(1 + SINR\right) \tag{15}$$

## III. Single User Problem

In this section, we focus on a CRN with just one SU and formulate an optimization problem to maximize its throughput. Since the log function is an increasing function of its argument; therefore the SU's rate maximization could be interpreted as is its SINR maximization. Considering equations 10-14, an optimization problem can be formulated in order to maximize the SU's throughput while the interference to PU's receiver is kept under a predetermined level based on PU's offered quality of service. The optimization problem is derived in equation (16).

$$\begin{aligned} &\underset{w_1, w_2}{Max}\ SINR \\ &s.t \begin{cases} P_{Int} \le I \\ \left\|\sqrt{P_2}\, w_1\right\|^2 \le P_2 \end{cases} \end{aligned} \tag{16}$$

where $I$ is the maximum acceptable level of interference to the PU's receiver.

Lemma 1: The optimization problem (16) can be robustly designed based on NBE model as:

$$\begin{aligned} &\underset{w_1, w_2}{Max}\ \frac{w_2{}^H (P_2\,(H_S\ w_1 w^H{}_1 H_S{}^H))w_2}{w_2{}^H (P_1\, h'_0 h'_0{}^H + P_1\,(2\sqrt{e}\,s\,\left\|h'_0\right\| + es^2) + s^2{}_{Noise} I_{N_r \times N_r})w_2} \\ &s.t \begin{cases} P_2\ w_1{}^H (h_0 h_0{}^H + (2\sqrt{e}\,s\,\left\|h_0\right\| + es^2) \times I_{n_t \times n_t})w_1 \le I \\ \left\|W_1\right\|^2 \le 1 \end{cases} \end{aligned} \tag{17}$$

Proof: The proof is given in appendix A-2.

Since receiver beamforming vector only appears in the objective function, it is applicable to firstly maximize objective function due to this vector.

$$\underset{w_1,w_2}{Max}\left(SINR\right) = \underset{w_1}{Max}\left(\underset{w_2}{Max}\left(SINR\right)\right) \tag{18}$$

Lemma 2: The optimization problem (17) can be derived based on receiver beamforming vector independent to the transmitter beamforming vector as:

$$\underset{w_2}{Max}\left(SINR\right) = l_{MAX}\left(\left((P_1\, h^{'}_0 h^{'}_0{}^H + P_1\,(2\sqrt{e}s\,\left\|h^{'}_0\right\| + es^{\,2}) + s^{\,2}{}_{Noise} I_{N_r \times N_r}\,)\right)^{-1} \times P_2\,(\,H_S\; w_1 w^H{}_1 H_S{}^H\,)\right) \tag{19}$$

Proof: The proof is presented in appendix A-3.

It is obvious that square matrix " $w_1 w^H{}_1$ ", is rank one. To further simplify the problem, the following three theorems are applied.

First, if A is a ranked-one matrix and B is a full rank matrix, the rank of the product matrix AB or BA will be one [14]. Second, if $l_1, l_2 ... l_N$ are Eigen values of A, the following relation is always true [14].

$$tr(A) = \sum_i l_i(A) = l_1(A) + l_2(A) + ... \tag{20}$$

Third, if A is ranked-one then it is straightforward to derive the following relation [14]:

$$tr(A) = l_{MAX}(A) \tag{21}$$

Thanks to the help of the above theorems, (19) can be restated as.

$$\underset{w_2}{Max}\left(SINR\right) = tr\left(\left((P_2\, h^{'}_0 h^{'}_0{}^H + P_2\,(2\sqrt{e}s\,\left\|h^{'}_0\right\| + es^{\,2}) + s^{\,2}{}_{Noise})\right)^{-1} \times P_1\,(\,H_S\; w_1 w^H{}_1 H_S{}^H\,)\right) \tag{22}$$

And then by use of the relation trace (AB) = trace (BA), it easily reformulated to:

$$\underset{w_2}{Max}\left(SINR\right) = tr\left(P_1 \times H_S{}^H \left((P_2\, h^{'}_0 h^{'}_0{}^H + P_2\,(2\sqrt{e}s\,\left\|h^{'}_0\right\| + es^{\,2}) + s^{\,2}{}_{Noise})\right)^{-1} \times (\,H_S\,) \times w_1 w^H{}_1\right) \tag{23}$$

It is worth noting that considering the following auxiliary parameters helps to more simplification.

$$A = P_1 \times H_S{}^H \left( (P_2\, h'_0 h'_0{}^H + P_2\, (2\sqrt{e}s\, \|h'_0\| + es^{\,2}) + s^{\,2}{}_{noise}) \right)^{-1} \times (\, H_S\,)$$
$$B = (h_0 h_0{}^H + (2\sqrt{e}s\, \|h_0\| + es^{\,2}) \times I_{n_t \times n_t}\,) \quad (24)$$

Applying these parameters leads to a simple form optimization problem which would be efficiently solved.

$$\underset{w_1}{Max} \quad tr(A \times w_1 w^H{}_1)$$
$$s.t \begin{cases} tr(B \times w_1 w^H{}_1) \le \dfrac{I}{P_2} \\ tr(w_1 w^H{}_1) \le 1 \end{cases} \quad (25)$$

In the remaining parts of this section; two methods are proposed to efficiently solve the above problem. The first solution utilizes SDP relaxation method to iteratively solve the problem and after all a close form solution is provided.

**A-SDP relaxation:**

Although the objective function and conditions of (25) are quadratic, but it is not a conventional quadratic constrained quadratic program (QCQP) problem because it minimize a convex objective function due to convex constraints. Since the objective and constraint functions in this paper are convex functions and maximizing a convex function subject to a convex constraint is not a trivial work; to efficiently solve the optimization problem, a SDP relaxation has been applied. As it is previously mentioned, the $w_1 w_1{}^H$ matrix is ranked-one. Suppose that $W == w_1 w_1{}^H$ . It is worth noting that the rank of the square matrix $W$ is 1. Hence, the optimization problem (25) can be simplified as:

$$\underset{W}{Max} \quad trace(AW)$$
$$s.t \begin{cases} trace(BW) \le \dfrac{I}{P_2} \\ trace(W) \le 1 \\ W = W^H \\ Rank(W) = 1 \end{cases} \quad (26)$$

While this problem has a simple linear programming form; the last constraint makes the problem non-convex, and therefore a huge computational burden is imposed by solving this optimization problem. To deal with this, the last constraint is discarded which is so called SDP relaxation form. The remaining problem is a linear programming which can be easily and effectively solved. Although discarding rank 1 constraint usually leads the optimization problem not to be optimal but as illustrates in [16], these kinds of problems always has a rank-1 solution. Simulation results also validate this simplification. The desired SDP relaxation based optimization problem is:

$$\begin{aligned} &\underset{W}{Max} \quad trace(AW) \\ &s.t \begin{cases} trace(BW) \le \dfrac{I}{P_2} \\ trace(W) \le 1 \\ W = W^H \end{cases} \end{aligned} \tag{27}$$

**B-Close form solution:**

In this subsection, a closed form solution is proposed for a specific case, where the same number of antennas existed for both the transmitter and receiver of secondary connection. In this case, (25) is simplified using trace (AB) = trace (BA) as:

$$\begin{aligned} &\underset{w_1}{Max} \quad w^H{}_1 \times A \times w_1 \\ &s.t \begin{cases} w^H{}_1 \times B \times w_1 \le \dfrac{I}{P_2} \\ w^H{}_1 \times w_1 \le 1 \end{cases} \end{aligned} \tag{28}$$

Theorem 1: The solution of the derived optimization problem is:

$$w_{1OPT} = \sqrt{\frac{I}{P_2}} \times \left\{ \left( B^{1/2} \right)^{-1} \times EigenVector \left( l_{MAX} \left( \left( B^{1/2} \right)^{-1H} A \left( B^{1/2} \right)^{-1} \right) \right) \right\} \tag{29}$$

Proof: The proof is presented in appendix A-4.

## IV. Generalized multi user problem

In the previous section, a closed form and a linear programming solution has been derived for CRN with only one pair of PU and SU transceiver operating concurrently at the same spectrum. In this section, an extended multiuser downlink network has been considered. Many conventional spectrum sharing techniques have been developed through recent years, e.g., CDMA and OFDMA. In this article, although the transmission scheme is not based on CDMA concept (spreading the spectrum of transmitting signal) because it needs specific cooperation among CR and PR systems; the basic idea of sharing a common spectrum for SUs without huge amount of cooperation between SUs-PUs has been introduced. On the other hand, the simple idea of splitting a spectrum band in to sub-bands in comparison with OFDMA, providing an interesting tool to construct a resource sharing system for a network of SUs. Regarding limited cooperation between SUs-PUs and the capability of directional transmission, a multiuser CR system using space-frequency holes is described in this section. Assume that there are $N_{SEC}$ receivers in in CR network in downlink transmission (Fig.2.). A general framework as an optimization problem for the two following cases is considered where there is only one base station which serves all SU's demand.

- Case 1: SU base station simultaneously sends SU's data in non-overlapping sub-bands of PU's spectrum and control the total interference power that is caused at PU receiver. See Fig.3.a.
- Case 2: SU base station sends SU's data in the whole spectrum of PU. See Fig.3.b.

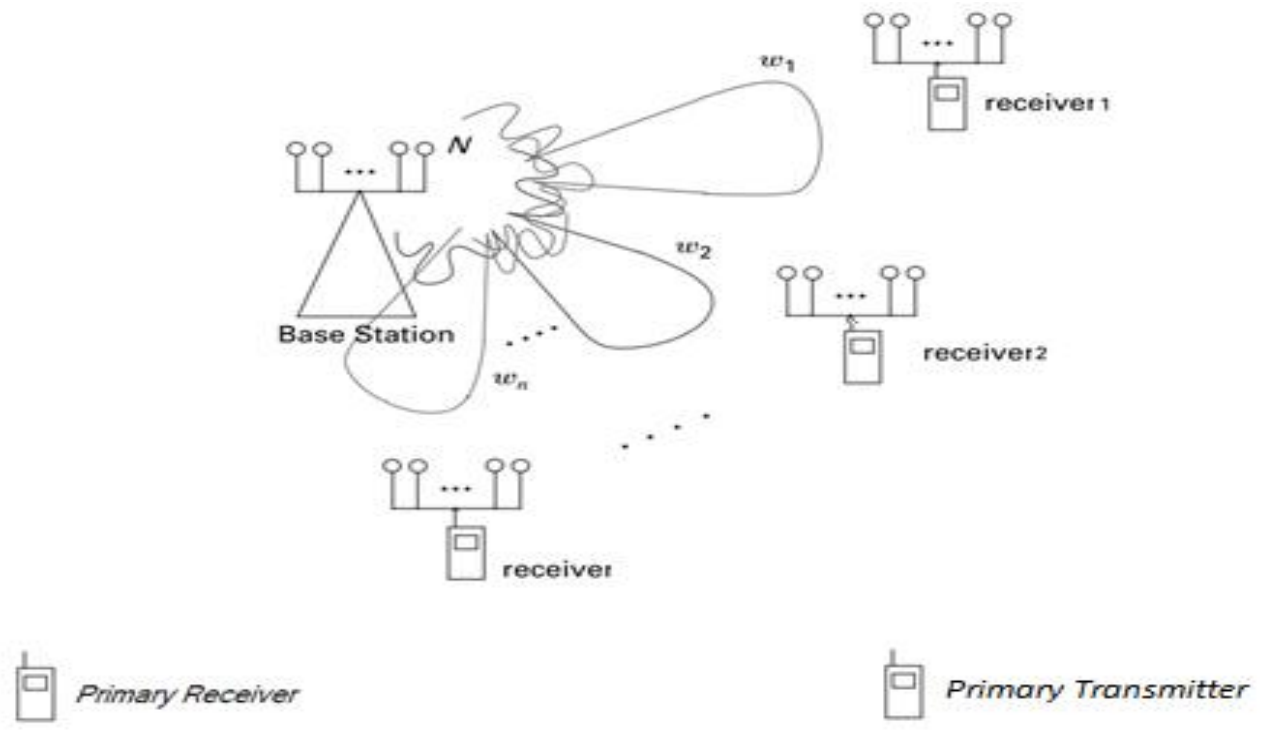

Fig.2. Multiuser transmission [7]

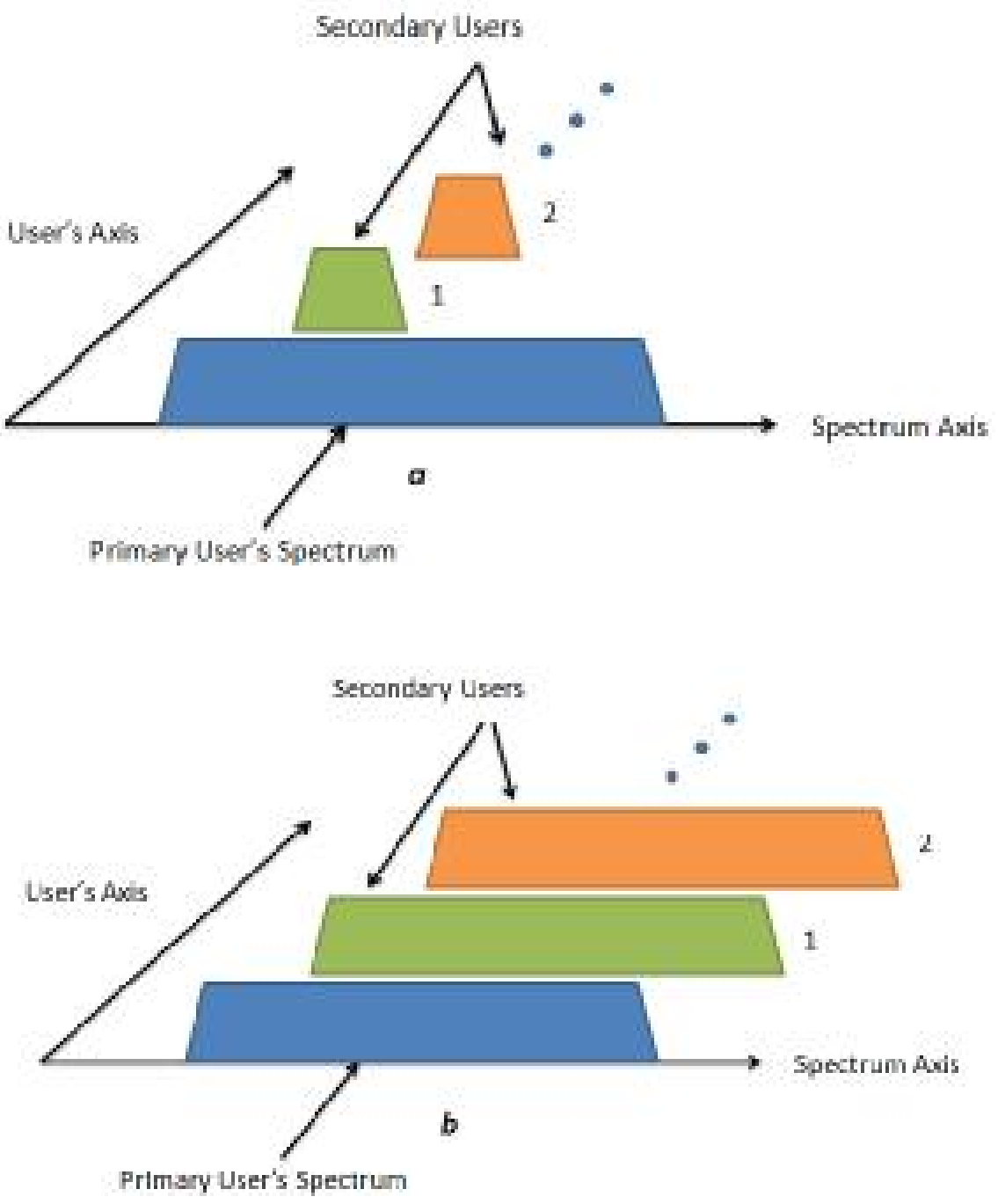

Fig.3. Transmission schemes for multiuser case

Moreover, the objective functions for these cases are defined as the sum rate of all SUs as:

$$\begin{aligned}
&Case\,1: Rate = \frac{1}{N_{SEC}} \sum_{K=1}^{N_{SEC}} \log_2\left(1 + SINR_K{}^1\right) \\
&Case\,2: Rate = \sum_{K=1}^{N_{SEC}} \log_2\left(1 + SINR_K{}^1\right)
\end{aligned} \qquad (30)$$

where:

$$\begin{aligned}
&SINR_K{}^1 = \frac{P_2 \left\| w_{1K}{}^H H_{SK} w_{2K} \right\|_2}{s^2{}_{Noise} + P_1 \left| w_{2K}{}^H h'_K \right|^2} \\
&SINR_K{}^2 = \frac{P_2 \left\| w_{1K}{}^H H_{SK} w_{2K} \right\|_2}{s^2{}_{Noise} + P_1 \left| w_{2K}{}^H h'_K \right|^2 + P_2 \sum_{\substack{J=1 \\ J \neq K}}^{N_{SEC}} \left\| w_{1J}{}^H H_{SJ} w_{2J} \right\|_2}
\end{aligned} \qquad (31)$$

**Case 1:**

In this case, the multiuser beamforming design in orthogonal sub-bands for CR system has been taken into account. The design parameters are beamforming vectors at both SU transmitter and receivers which should be derived in a way to maximize the sum rate (throughput) of CRN. Similar to the previous section, $H_{SK}$ is the channel matrix between SU transmitter and k<sup>th</sup>-SU receiver. Furthermore, $h / h'_K$ is the channel vector between SU/PU transmitter and PU/k<sup>th</sup> -SU receiver. $w_{1K}$ and $w_{2K}$ vectors are also transmitter/receiver beamforming vectors related to k<sup>th</sup> SU. Therefore:

$$\begin{aligned} &\underset{w_{1K} \& w_{2K}}{Max} \left( \frac{1}{N_{SEC}} \sum_{K=1}^{N_{SEC}} \log_2 \left( 1 + SINR_K{}^1 \right) \right) \\ &s.t \;\; P_2 \sum_{K=1}^{N_{SEC}} \left| w_{1K}{}^H h \right|^2 \leq I \end{aligned} \tag{32}$$

Since there is no way to guaranty short term fairness among SU's in equation (32); In the following a fairness based design is also proposed which its performance is near optimal. In addition, it is easy to show that the constraint in equation (32) would become in to equality. To simplify the problem without loss of generality, a middle stage solution has been considered in order to efficiently solve the problem. First, consider that the US transmitter serves each SU's receiver with interference power limit $I^K$ where obviously $\sum_{K=1}^{N_{SEC}} I^K = I$ is always guarantied. Therefore, the optimization problem (32) is converted into the following sub-problem:

$$\begin{aligned} &\underset{w_{1K} \& w_{2K}}{Max} \;\; SINR_K{}^1 \\ &s.t \;\; P_2 \left| w_{1K}{}^H h \right|^2 = I^K \end{aligned} \tag{33}$$

In previous section, it has been proofed that the close form solution for problem (33) is as follows.

$$\underset{w_{1K} \& w_{2K}}{Max} \;\; SINR_K{}^1 \propto I^K \rightarrow \underset{w_{1K} \& w_{2K}}{Max} \;\; SINR_K{}^1 = I^K \times y^{\,K} \tag{34}$$

Where, the $y^{K}$ parameter is the solution of the equation (29) for $I^{K} = 1$ which has a meaningful interpretation. The better channel conditions the higher level of $y^{K}$ is available. This middle stage solution takes a major role in the design of the whole network in equation (32) which leads the original problem to be converted to the following problem.

$$\begin{aligned} &\underset{I_K}{Max}\left(\sum_{K=1}^{N_{SEC}} \log_2\left(1 + I^{K} \times y^{K}\right)\right) \\ &s.t \sum_{K=1}^{N_{SEC}} I^{K} = I \end{aligned} \tag{35}$$

Lemma 3: The optimal close form solution for the optimization problem in equation (35) is:

$$I^{J} = \frac{1}{N_{SEC}}\left(I + \sum_{K=1}^{N_{SEC}} \frac{1}{y^{K}}\right) - \frac{1}{y^{J}} \tag{36}$$

Proof: See appendix A-5.

This solution implies that, the higher level of the $y^{K}$ ,the more interference could SU transmitter cause to the PU receiver. Although this case maximizes the sum rate of the SU network but it is somehow a greedy method therefore; short term fairness would not be guaranteed. The algorithm of how to calculate the optimum solution is described in table 2. In order to provide some sorts of fairness between SU's receivers, the following simple relation between the design parameters is assumed.

$$\log_2\left(1 + I^{K} \times y^{K}\right) = \log_2\left(1 + I^{J} \times y^{J}\right) \tag{37}$$

Which it means that, each SU's rate is considered to be the same as others therefore; short term and long term fairness would be promised.

Lemma 4: The close form solution for fairness based design is as:

$$I^J = \frac{\frac{I}{y^J}}{\sum_{J=1}^{N_{SEC}} \frac{1}{y^J}} \tag{38}$$

Proof: See appendix A-6.

In contrast with the non-fairness based design, the larger the $y^K$ the smaller interference limit will be available. This simple method controls the greedy behavior of SU's rate allocation. In the simulation section, it has been shown that this method not only provides fairness but also is near optimal.

**Case 2:**

In this case, SU base station simultaneously sends each SU's data in the whole spectrum. SU transmitter should design specific and different transmitter beamforming vectors $w_{1k}$ for each secondary receiver in order to maximize the throughput of CR network. In this case, an optimization problem based on second objective function in equation (31) has been evaluated as:

$$\begin{aligned} &\underset{w_{1K} \& w_{2K}}{Max} \left( \sum_{K=1}^{N_{SEC}} \log_2 \left(1 + SINR_K{}^2\right) \right) \\ &s.t \;\; P_2 \sum_{K=1}^{N_{SEC}} \left| w_{1K}{}^H h \right|^2 \le I \end{aligned} \tag{39}$$

It is worth noting that each SU's data should be considered as interference to both PU receiver and other SU receivers. Since this problem is very complicated; another schemes which simplify the design are proposed in the rest of this section. Obviously, the objective function or equivalently SINR is not a convex function of beamforming vectors; therefore the exhaustive search to find the optimal solution would be a very time consuming process (if there is a feasible solution). Two sub-optimal solutions to overcome these difficulties are described here.

1- Beamforming vectors are derived based on the solution presented in case 1. In this case, the inter-system interference between SUs has not taken into account therefore the solution would be sub optimal. In the simulation section it has been discussed that the performance of this case is far from optimal but it is suggested only for its simple close form solution.

2- Beamforming design is based on intra-system interference control scheme. This case applies a limit on interference power exerted by each SU on other SUs which is described below.

**Interference Control Solution:**

A simple but efficient way to solve the network beamforming design is proposed based on considering that the SU transmitter not only controls the amount of interference at PU receiver but also handles inter-system interference between SUs. Interference control parameter is shown as $I^{'}$ in equation (40). Therefore, the complicated objective function in equation (39) is converted into:

$$\begin{aligned} &\underset{w_{1K}\,\&\,w_{2K}}{Max}\left(\sum_{K=1}^{N_{SEC}}\log_2\left(1+SINR_K{}^1\right)\right) \\ &s.t\;\begin{cases} P_2\sum_{K=1}^{N_{SEC}}\left|w_{1K}{}^H h\right|^2 \le I \\ Interference\ to\ Other\ SUs \le I^{'} \end{cases} \end{aligned} \tag{40}$$

Moreover, to simplify the problem it is assumed that individual SU’s interference limit on PU’s receiver, rely on $I^K \le \frac{I}{N_{EC}}$. A simple manipulation shows:

$$\sum_{K=1}^{N_{SEC}} I^K \le I \tag{41}$$

It is easily verifiable to show the simplified optimization problem would be converted to:

$$\begin{aligned} &\underset{w_{1K}\,\&\,w_{2K}}{Max}\left(SINR_K{}^1\right) \\ &s.t\;\begin{cases} P_2\left|w_{1K}{}^H h\right|^2 \le \frac{I}{N_{SEC}} \\ Interference\ to\ Other\ SUs \le I^{'} \end{cases} \end{aligned} \tag{42}$$

This would be the simplest form of the optimization problem which the objective function is convex function of transmitter beamforming vectors. In addition, the conditions also are convex therefore problem (42) could be solved easily by SDP relaxation method which has been described in section 2.

| Multiuser Design Algorithm: Case 1 |
|---|
| 1. For each SU derive $y^K$<br>2. Use the optimum interference design (36) to calculate $I^K$<br>3. If all $I^K$ are positive, go to the next step, otherwise omit the user that has negative $I^K$ from transmission list and run the procedure again. Those which has negative $I^K$ should not be considered to receive data.<br>4. Use $I^K$ and derive the optimum throughput of the network<br>5. end |

## V. Simulation result

In this section, numerical results are presented to evaluate the effectiveness of the proposed methods. Since the SU and PU system has different physical layer, in order to describe them specifically the table 1 provide the simulation setup. It is assumed that the total amount of antennas available at both SU's transmitter-receiver is 10. Since the performance of SU network clearly depends on the amount of available antennas and further; SU pair simultaneously should control the mutual interference between the two networks, therefore it has been assumed that each secondary TX/RX has 5 antennas. It easy to proof that if NBE model for channels between two networks are the same; the optimum number of antennas (for the highest SU's bit rate per interference exerted on PU receiver) at SU's TX/RX is 5. Also, each PU is considered to have Omni-directional antenna. The power budget for both primary/secondary TX's are 20 db in compare to noise power (noise power assumed to be 0 db). Maximum cumulative interference limit for primary receiver is also considered to be 5 db. Furthermore, the number of secondary users in the multiuser SU design is 3. It is obvious that, the secondary TX can totally eliminate the interference power to the primary receiver in no-error case; thus we only consider errors greater than 0. The channel parameter $s$ is assumed to be 1.

The results depicted in this paper are based on the average of 1000×1000 simulation runs using different channel realizations and channel error in NBE model. To be fair in comparison, the proposed methods results and the previous works results in single user case will be discussed.

| Parameters | Amount |
|---|---|
| No TX/RX Antenna for Secondary System | $N_r$ , $N_t \leq 9$ , $N_t + N_t = 10$ |
| No TX/RX Antenna for Primary System | 1 |
| Maximum TX Power for Secondary System (normalize to noise power) | 20 db |
| Maximum TX Power for Secondary System (normalize to noise power) | 20 db |
| Maximum Interference Power to Primary RX | 5 db |
| No Secondary Users for Multi User Case | 3 |

Table 1: Simulation Setup

Single user:

Fig.4. shows the single user CRN's performance over the interference limit. Obviously, the more interference limit available, the more SINR and consequently the more rates are available. This figure also shows how the performance of SU degrades with grows of the channel error amount. It shows that the robust design is very sensitive to the channel error between two networks. For errors greater than 3, the SU's performance degrades catastrophically because for such a large amount of error, both SU transmitter and receiver could not estimate and control the direction of the signal which considered as interference. Therefore for large errors, beamforming may not be very effective enough.

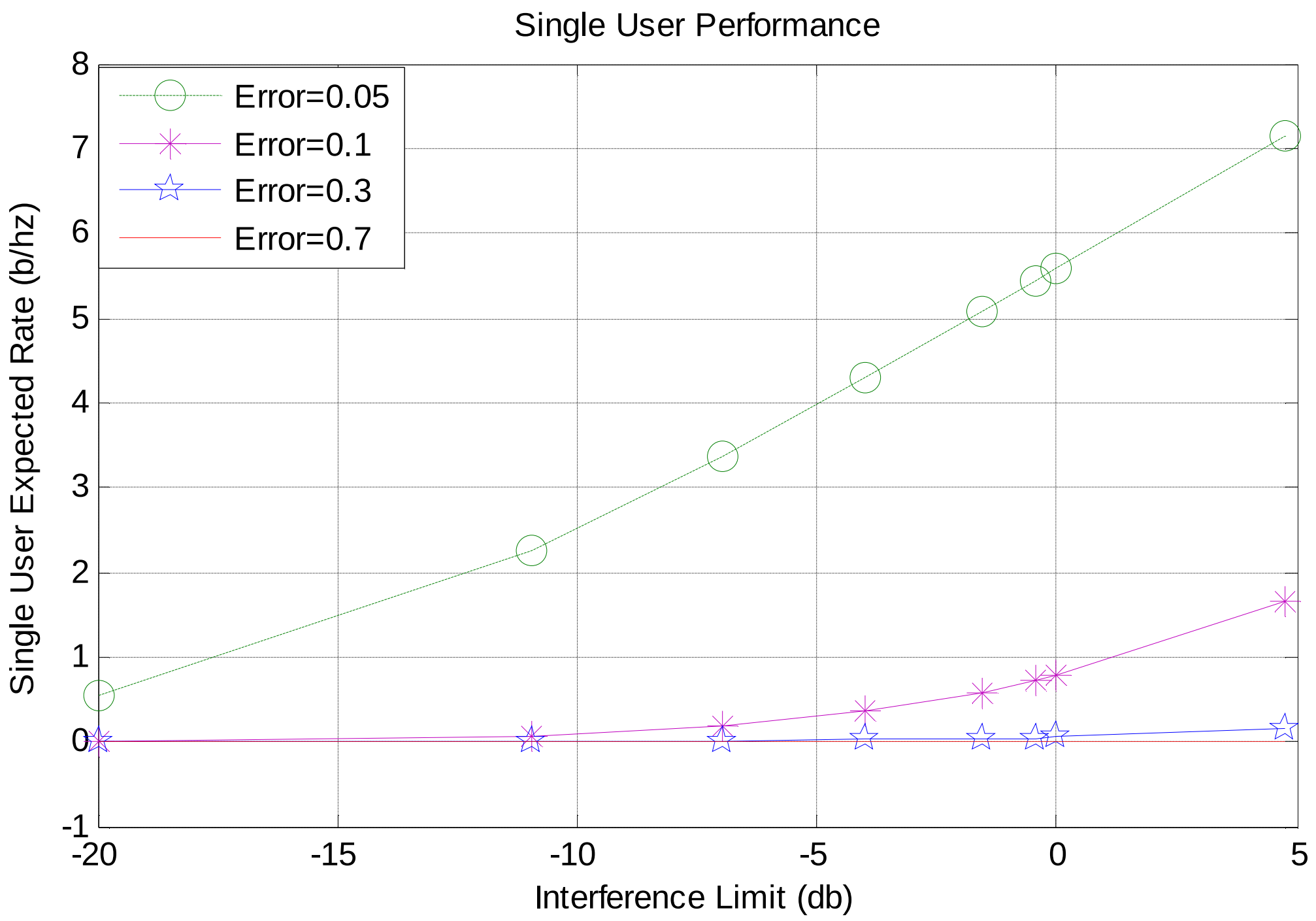

Fig.4. Single user performance of SU over interference limit

Moreover, fig.5. shows the interference power at the PU’s receiver which completely depends the interference limit. As equation (16) demonstrates, it is obvious that the interference at the PU’ receiver is always lower than the interference limit because the beamforming design is based on the worst case therefore in the majority of cases, the inequality is guaranteed. Fig.6. shows the SU’s rate per channel error for fixed interference limit. As mentioned before, it is obvious that the performance of CR system is a very sensitive function of channel error. Since for lower amount of error, it is applicable to control the amount of intra system interference so the slope of the diagram is very sharp but for larger amount of errors neither the SU transmitter nor the receiver are capable of controlling the interference so the slope would be very smooth.

Furthermore, fig.7. Describes the interference power which the SU transmitter exerts on PU receiver. Obviously channel error degrades the system performance and in this case, the large interference at PU receiver is the effect of the large amount of channel error.

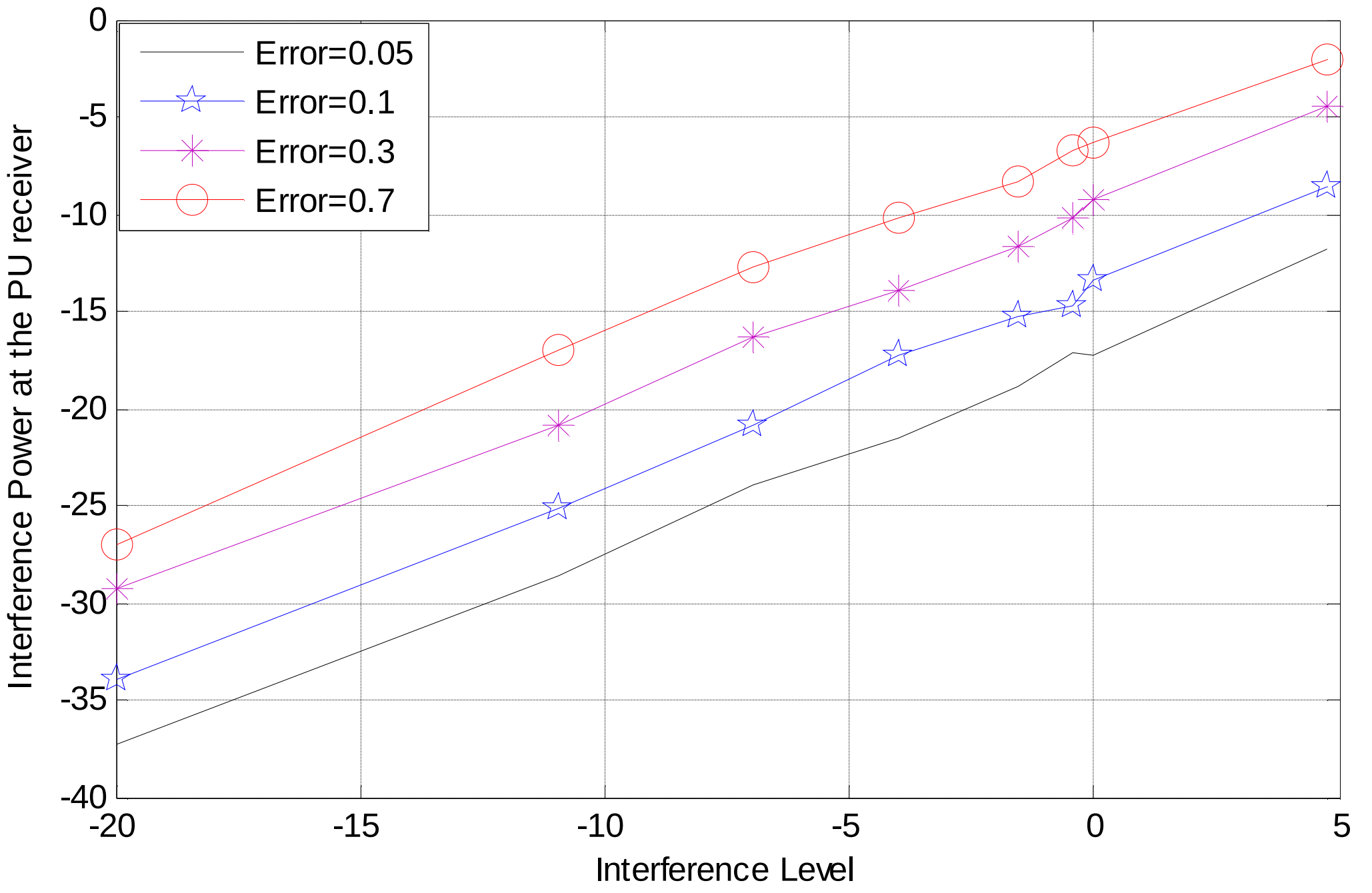

Fig.5. Interference Power at the PU receiver

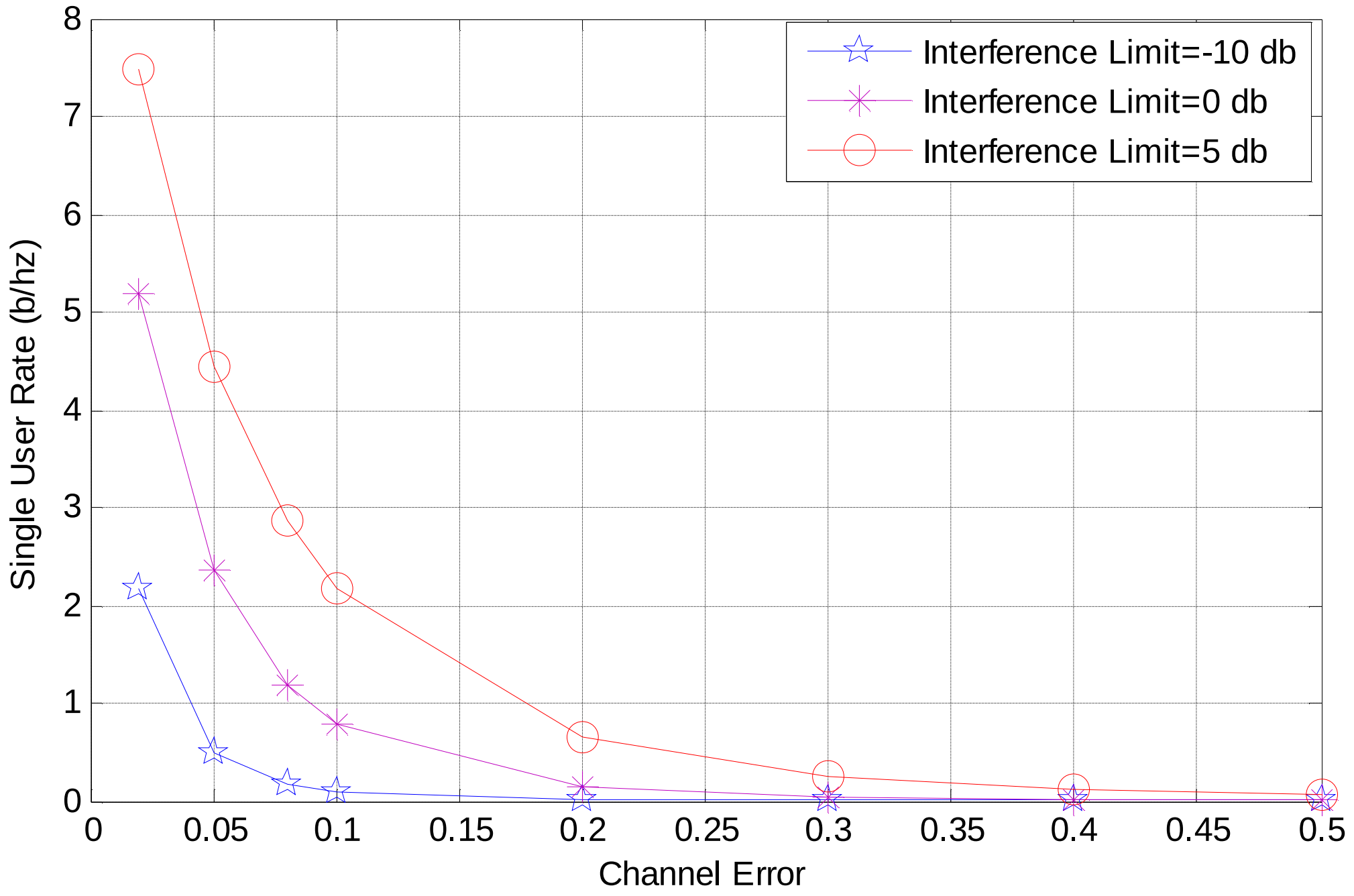

Fig.6. SU's performance over channel error

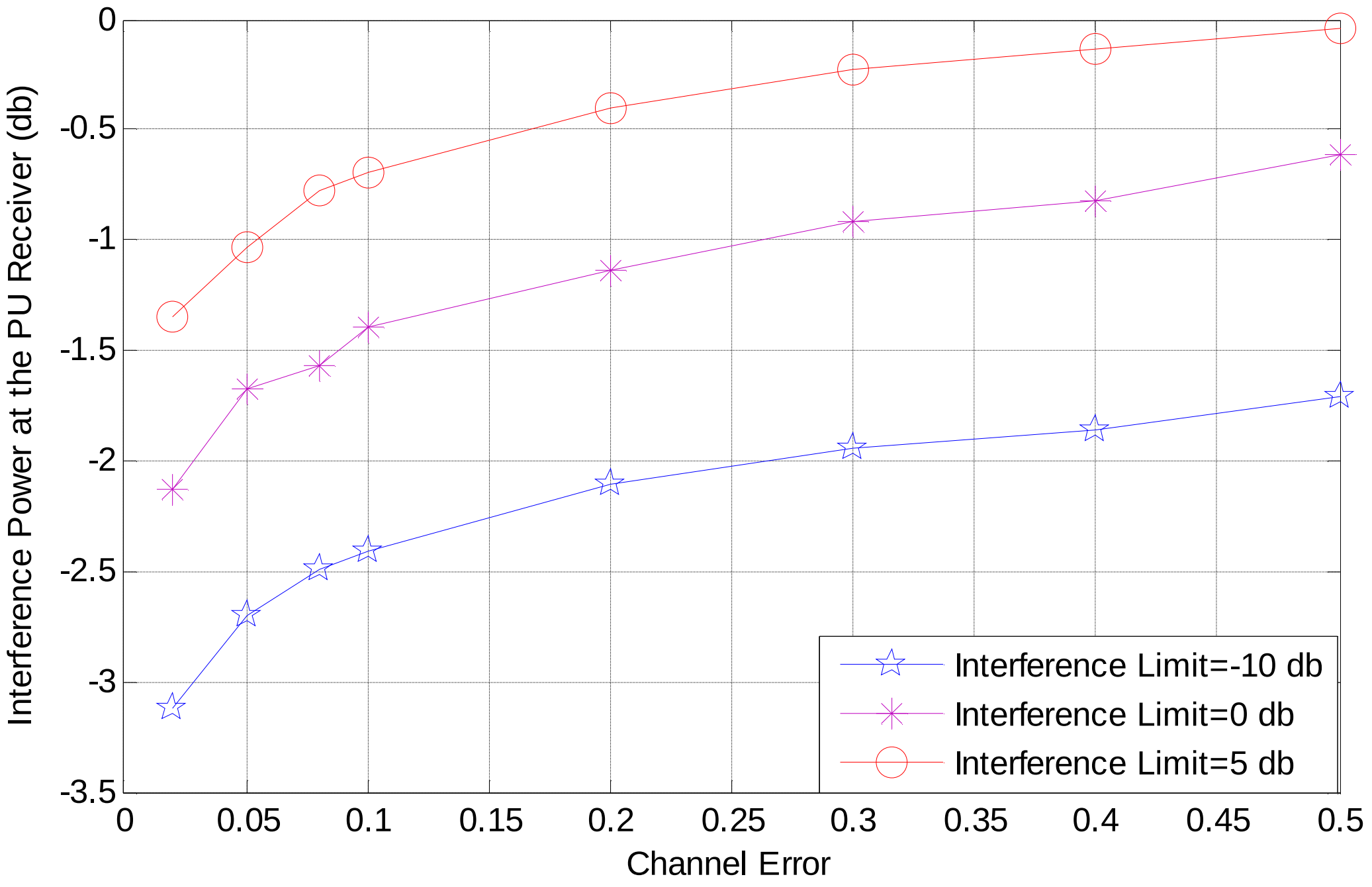

Fig.7. Interference power at PU over channel error

Multiuser case:

In the multiuser design, two general design using OFDM and UWB concept has been proposed. Fig.8. , Fig.9. Shows the performance of the OFDM based design considering fairness. As mentioned in the text, the two fair and non-fair designs are little bit different in their performances. It shows that the non-fair design has as good performance as fair design with a more capability of guaranteeing the short-term fairness. Both methods guarantee the long-term fairness because of the averaging of their performances over all transmission times. In Fig.10, a comparison of short-term fairness between two design approaches is presented. In consecutive time slots, the fairness-based design for all SUs provides a complete fairness while the non-fairness based design has no such a guarantee. In this figure, SU's rate diagrams are based on the design parameter $I$=1 (interference limit).

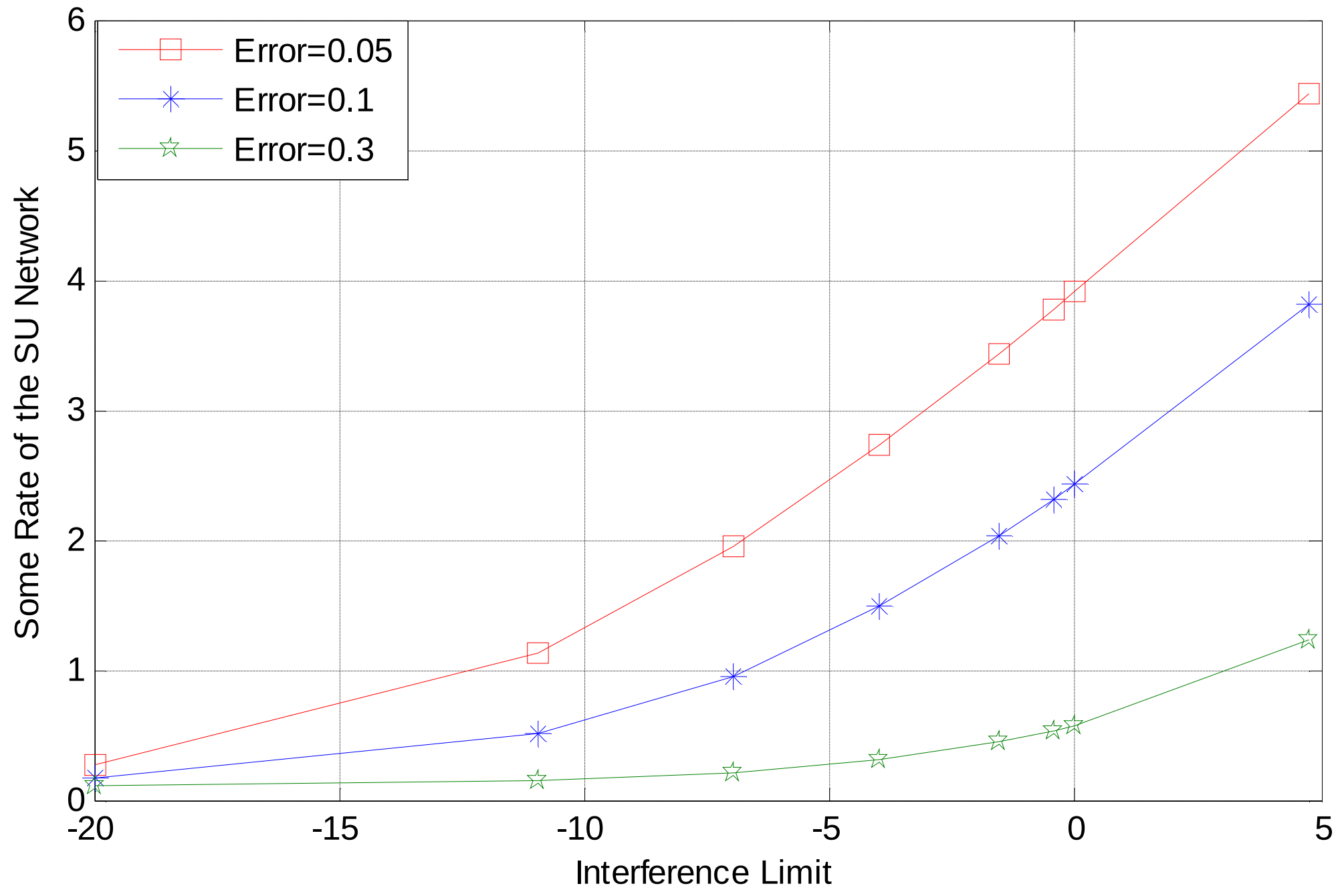

Fig.8. Multiuser OFDM beamforming performance with fairness

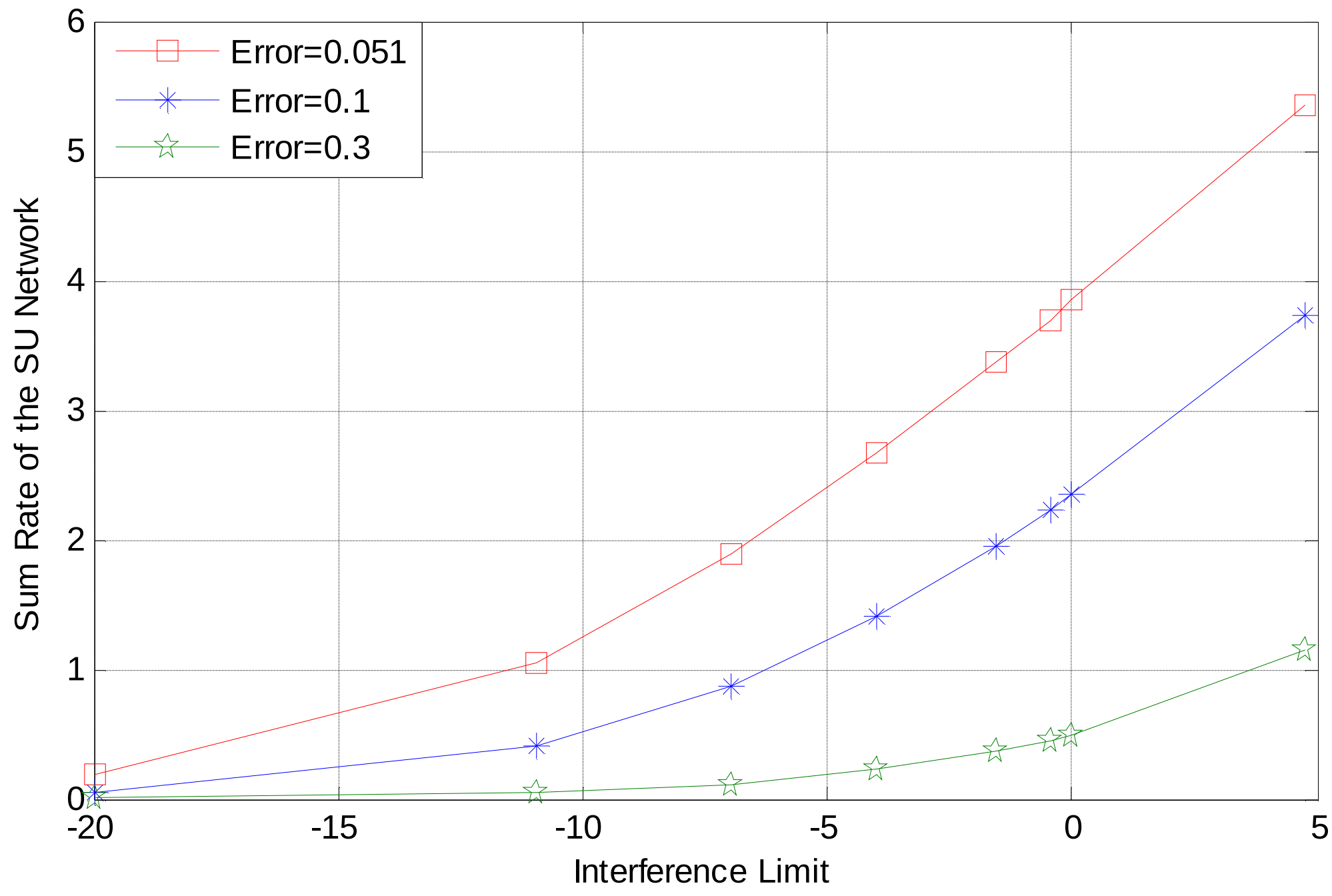

Fig.9. Multiuser OFDM beamforming performance without fairness

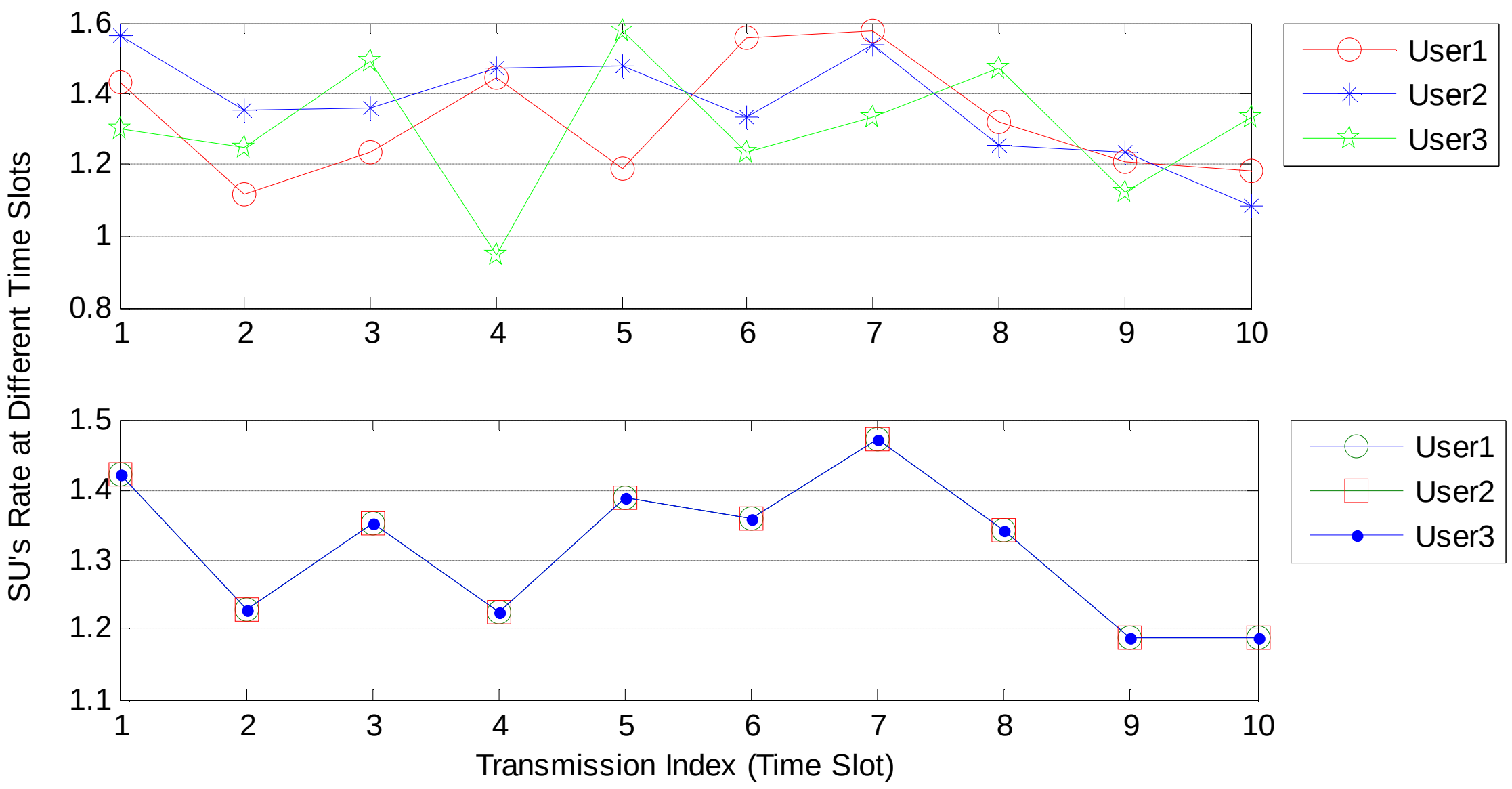

Fig.10. Comparison between fair and non-fair based design

Fig.11. shows the performance,

## VI. Discussion

This article provided new and effective design methods which consider MIMO beamforming considering partial channel state information. The robust optimum/suboptimum beamforming design for MIMO cognitive pair/network has been proposed. It is assumed that the CSI between all secondary users is known completely but CSI between primary and secondary users is partially known. Proposed methods consider this imperfect knowledge of channels and derive simple closed-form solution in most of the cases. For single user design, uniform distribution of antennas between SU transmitter and receiver enables CR system to overcome destructive intra system interference with PU network. This selection of antenna, has guaranteed that the performance of proposed methods are better than previous works design. Furthermore, for multi user CRN, two concepts have been considered. OFDM based design leads to a closed form solution for both considering fairness or not. The idea of UWB and the capacity (rate) improvement also has taken into account which leads to an outstanding performance. Finally, it is worth noting that all these proposed methods could be generalized for every networks coexisted with each other. The more precious CSI leads to a remarkable improvement of spectrum scarcity.

**Appendix:**

Appendix 1:

Assume that $x$ is an n component random vector as $x : N(m,P)$ which the entropy of this vector could be shown as $H(x) = n\log_2(2pe) + \log_2(|P|)$ [15]. For MIMO system which we have $y = Hx + n$, it has been shown that if the channel matrix H assumed to be precisely known, then the maximum mutual information between $(y,x)$ that equals the channel capacity is as [15]:

$$C = MAX(M(y,x|H)) = \log_2(\left|\frac{1}{s^2}HPH^H + I\right|)$$

Where $P$ is the covariance matrix of the input vector $x$ and $s^2$ is noise power. Furthermore, the transmission beamforming model for CR system is as:

$$y = Hx + h'z + n$$

Where $z$ is the PU's data symbol. Similarly, the mutual information between $x$ and $y$ could be easily derived. We assume that all channels matrixes are known.

$$M(y,x|H,h') = M(y|H,h') - M(y|x,H,h')$$

To maximize the above formula, as discussed in [15], all the SU's data should have a Gaussian distribution. Similarly we have:

$$\begin{cases} Cov(y|H,h') = HPH^H + p'h'h'^H + s^2 I \\ Cov\left(y|x,H,h'\right) = p'h'h'^H + s^2 \end{cases}$$

Where $P$ and $p'$ are subsequently SU's covariance matrix of transmitted signal and PU's signal power. By inserting this formula to the mutual information equation, the capacity could be shown as:

$$C = \log_2\left(\left|HPH^H + p'h'h'^H + s^2 I\right|\right) - \log_2\left(\left|p'h'h'^H + s^2 I\right|\right) = \log_2\left(\left|I + \frac{HPH^H}{p'h'h'^H + s^2 I}\right|\right)$$

Similarly, by considering the jointly transmission-reception beamforming at CR system, the model would be modified as:

$$y_i = x_i \times w_2^H \times H_s \times w_1 + z_i \times w_2^H \times h' + w_2^H \times n_i$$

It is can be easily verified that the capacity is as following:

$$C = MAX\left(M(y, x | H, h')\right) = \log_2\left(1 + \frac{w^H{}_2\left(pHw_1 w^H{}_1 H^H\right)w_2}{w^H{}_2\left(p'h'h'^H + s^2 I\right)w_2}\right) = \log_2\left(1 + SINR\right)$$

Appendix 2:

Regarding the robust design concept, the design vectors should operate well specially in the worst channel condition. Due to the channel error model, the following relations could be easily verified:

$$\begin{aligned} &\left|h - h_0\right| \le \sqrt{e}\, s \;\rightarrow\; h = h_0 + \sqrt{e}\, s \times u \qquad where \quad |u| \le 1 \\ &\left|h' - h'_0\right| \le \sqrt{e}\, s \;\rightarrow\; h' = h'_0 + \sqrt{e}\, s \times u' \qquad where \quad |u'| \le 1 \end{aligned}$$

We use these interpretations and solve the beamforming problem in the worst case CSI. First, we insert the above forms in the objective function.

$$\underset{CSI\ Error}{Min}\left(SINR\right) = \frac{P_2 \times \left|w_2{}^H \times H_s \times w_1\right|^2}{\underset{CSI\ Error}{Max}\left(P_1 \times \left|w_2{}^H \times h'\right|^2\right) + s^2{}_{noise} \times w_2{}^H \times w_2}$$

Then we have:

$$\begin{cases} \underset{\left|h' - h'_0\right| \le \sqrt{e}\, s}{Max}\left(P_1\left|w_2{}^H h'\right|^2\right) \\ \left|h' - h'_0\right| \le \sqrt{e}\, s \;\rightarrow\; h' = h'_0 + \sqrt{e}\, s \times u' \qquad where \quad |u'| \le 1 \end{cases}$$

For further simplifications, we consider:

$$\begin{cases} \left|W_2^H h^{'}\right|^2 = W_2^H h^{'} h^{'H} W_2 \\ h^{'} h^{'H} = (h_0^{'} + \sqrt{es} \times u')^H (h_0^{'} + \sqrt{es} \times u') \end{cases}$$

where:

$$h^{'} h^{'H} = (h_0^{'} + \sqrt{es} \times u')^H (h_0^{'} + \sqrt{es} \times u') = h_0^{'H} h_0^{'} + \Delta$$
$$which \ \Delta = \sqrt{es}\,(h_0^{'H} u' + h_0^{'} u'^H) + es^2 \times u'^H u'$$

It is obvious that $\Delta$ is norm bounded as:

$$\|\Delta\| = \left\|\sqrt{es}\,(h_0^{'H} u' + h_0^{'} u'^H) + es^2 \times u'^H u'\right\| \le \sqrt{es}\,\left\|h_0^{'H} u'\right\| + \sqrt{es}\,\left\|h_0^{'} u'^H\right\| + es^2 \times \left\|u'^H u'\right\|$$

Which the $\|A+B\| \le \|A\| + \|B\|$ inequality has taken in to account. Also we use Cauchy-Schwartz inequality:

$$\|\Delta\| \le \sqrt{es}\,\left\|h_0^{'H}\right\| \|u'\| + \sqrt{es}\,\left\|h_0^{'}\right\| \left\|u'^H\right\| + es^2 \times \|u'\|^2$$

Or equivalently:

$$\|\Delta\| \le 2\sqrt{es}\,\left\|h_0^{'H}\right\| \|u'\| + es^2 \le 2\sqrt{es}\,\left\|h_0^{'H}\right\| + es^2$$

Therefore the worst case design could be shown as:

$$\underset{\left|h^{'} - h_0^{'}\right| \le \sqrt{es}}{Max} (P_1 \left|w_2^H h^{'}\right|^2) = \underset{\left|h^{'} - h_0^{'}\right| \le \sqrt{es}}{Max} (P_1\, w_2^H h^{'} h^{'H} w_2) = Max(P_1\, w_2^H (h_0^{'H} h_0^{'} + \Delta) w_2)$$

$$= P_1\, w_2^H h_0^{'H} h_0^{'} w_2 + Max(P_1\, w_2^H \Delta w_2)$$

Using Cauchy-Schwartz inequality once more, the following relation is derived:

$$Max(P_1\, w_2{}^H \Delta w_2) = P_1\, Max\, \left\| w_2{}^H \Delta w_2 \right\| \overset{Cauchy-Schwartz}{\leq} P_1\, Max \left\| w_2{}^H \right\| \left\| \Delta \right\| \left\| w_2 \right\| \leq P_1\, (2\sqrt{e}\, s\, \left\| h_0^{'H} \right\| + es^{\,2}) \left\| w_2 \right\|^2$$

Finally:

$$\underset{\left| h^{'} - h_0^{'} \right| \leq \sqrt{e}\, s}{Max} (P_1 \left| w_2{}^H h^{'} \right|^{\,2}) = P_1\, w_2{}^H h_0^{'H} h_0^{'} W_2 + P_1\, (2\sqrt{e}\, s\, \left\| h_0^{'H} \right\| + es^{\,2}) \left\| w_2 \right\|^2$$

Then the worst case SINR is as follows:

$$\underset{CSI\ Error}{Min} \left( SINR \right) = \frac{P_2 \times \left| w_2{}^H \times H_s \times w_1 \right|^2}{P_1\, w_2{}^H h_0^{'H} h_0^{'} w_2 + P_1\, (2\sqrt{e}\, s\, \left\| h_0^{'H} \right\| + es^{\,2}) \left\| w_2 \right\|^2 + s^{\,2}_{noise} \times w_2{}^H \times w_2}$$

Appendix 3:

Consider two square matrixes A and B. We want to maximize the following function by designing a vector *w*.

$$f(w) = \frac{w^H A w}{w^H B w}$$

We assume that the maximum of this function is as follows:

$$\underset{w}{MAX} \left( f(w) \right) = U$$

We know from [18], that the complex vector differentiation has the following properties.

$$\begin{cases} \dfrac{\partial \left( w^H A w \right)}{\partial w} = 2Aw \\ \dfrac{\partial \left( w^H A w \right)}{\partial w^H} = 0 \end{cases}$$

Then we differentiate the *f(w)* function to find critical vectors:

$$\frac{\partial(f(w))}{\partial w} = \frac{\partial\left(\frac{w^H Aw}{w^H Bw}\right)}{\partial w} = \frac{2Aw\left(w^H Bw\right) - 2Bw\left(w^H Aw\right)}{\left(w^H Bw\right)^2} = 0$$

Therefore:

$$Aw\left(w^H Bw\right) - Bw\left(w^H Aw\right) = 0 \rightarrow Aw = Bw\frac{\left(w^H Aw\right)}{\left(w^H Bw\right)}$$

We know that for column vector *w*, the *f(w)* function is only a real number.

$$\begin{cases} B^{-1}Aw = \dfrac{\left(w^H Aw\right)}{\left(w^H Bw\right)} w \\ f(w) = \dfrac{\left(w^H Aw\right)}{\left(w^H Bw\right)} \end{cases} \rightarrow \left(B^{-1}A\right) w = f(w) \times w$$

Where *f(w)* is the Eigen value of ($BA^{-1}$ ) matrix. Then if we want to maximize the objective function, it is obvious that it should be as follow:

$$\underset{w}{MAX}\left(f(w)\right) = U = l_{MAX}\left(B^{-1}A\right)$$

Appendix 4:

We have the following form to solve:

$$\begin{aligned}&\underset{w_1}{Max} \quad w^H{}_1 \times A \times w_1 \\ &s.t \begin{cases} w^H{}_1 \times B \times w_1 \le \dfrac{I}{P_2} \\ w^H{}_1 \times w_1 \le 1 \end{cases} \qquad and\ consider\ \frac{I}{P_2} = g\end{aligned}$$

First, we don't consider the last condition, and then we will check that the solution should guarantee this inequality. For further simplification, the following auxiliary parameter is assumed.

$$w_{Auxillary} = \left(B^{1/2}\right) w_1$$

Therefore the optimization problem is easily converted to the following:

$$\begin{aligned}&\underset{w_{Auxillary}}{Max} \quad w_{Auxillary}{}^{H} \left( \left(B^{1/2}\right)^{-1H} A \left(B^{1/2}\right)^{-1} \right) w_{Auxillary} \\ &s.t \ \ w_{Auxillary}{}^{H} w_{Auxillary} \le g\end{aligned}$$

We will prove that the optimal solution must satisfy the equality for the condition. We assume that in the optimal solution, the equality is not satisfied then we prove that this assumption is wrong. It is obvious that:

$$\frac{w_{OPTIMUM\ Auxillary}{}^{H} w_{OPTIMUM\ Auxillary}}{g} \le 1$$

We choose a number like J>1 that satisfy the following condition (it can be easily shown that such number would exist).

$$\frac{w_{OPTIMUM\ Auxillary}{}^{H} w_{OPTIMUM\ Auxillary}}{g} \le \frac{1}{J} \le 1 \rightarrow J\left(w_{OPTIMUM\ Auxillary}{}^{H} w_{OPTIMUM\ Auxillary}\right) \le g$$

We define another vector that satisfy inequality condition and has a grater objective function than the optimal solution, which contradicts with the assumption. Therefore the inequality must became to equality.

$$w' = \sqrt{J} \times w_{OPTIMUM\ Auxillary}$$

Clearly:

$$w'^{H}\, w' = J \times w_{OPTIMUM\ Auxillary}{}^{H} \times w_{OPTIMUM\ Auxillary} \leq g$$

Also by inserting this vector in the objective function:

$$w'^{H}\left(\left(B^{1/2}\right)^{-1H} A\left(B^{1/2}\right)^{-1}\right)w' = J \times w_{OPTIMUM\ Auxillary}{}^{H}\left(\left(B^{1/2}\right)^{-1H} A\left(B^{1/2}\right)^{-1}\right)w_{OPTIMUM\ Auxillary} >$$
$$w_{OPTIMUM\ Auxillary}{}^{H}\left(\left(B^{1/2}\right)^{-1H} A\left(B^{1/2}\right)^{-1}\right)w_{OPTIMUM\ Auxillary}$$

Which contradict with the optimality of $w_{OPTMUM}$. Now we solve the problem and find a close form solution. By the equality condition, we write the Lagrange function for the optimization problem.

$$f\left(l, w_{Auxillary}\right) = w_{Auxillary}{}^{H}\left(\left(B^{1/2}\right)^{-1H} A\left(B^{1/2}\right)^{-1}\right)w_{Auxillary} - l\left(w_{Auxillary}{}^{H} w_{Auxillary} - g\right)$$

We differentiate the Lagrange function:

$$\begin{cases} \dfrac{\partial f}{\partial l} = 0 \rightarrow w_{Auxillary}{}^{H} w_{Auxillary} = g \\ \dfrac{\partial f}{\partial w_{Auxillary}} = 0 \rightarrow \left(\left(B^{1/2}\right)^{-1H} A\left(B^{1/2}\right)^{-1}\right)w_{Auxillary} = l\, w_{Auxillary} \end{cases}$$

We know from the second equation:

$$l \text{ is eigen value of } \left(\left(B^{1/2}\right)^{-1H} A \left(B^{1/2}\right)^{-1}\right) matrix$$

And by multiplication of both sides with the optimal vector we have:

$$w_{Auxillary}{}^{H}\left(\left(B^{1/2}\right)^{-1H} A \left(B^{1/2}\right)^{-1}\right) w_{Auxillary} = l\, w_{Auxillary}{}^{H} w_{Auxillary} = l g$$

Then:

$$Max\left( w_{Auxillary}{}^{H}\left(\left(B^{1/2}\right)^{-1H} A \left(B^{1/2}\right)^{-1}\right) w_{Auxillary} \right) = g\, Max(\, l\,) = g \times l_{MAX}\left(\left(B^{1/2}\right)^{-1H} A \left(B^{1/2}\right)^{-1}\right)$$

Which means that:

$$w_{Auxillary} = b \times eigen\ vector\left( l_{MAX}\left(\left(B^{1/2}\right)^{-1H} A \left(B^{1/2}\right)^{-1}\right)\right)$$

$$w_{Auxillary}{}^{H} w_{Auxillary} = g = b^{2}$$

By obtaining the optimum beam forming vector, from auxiliary vector, we have:

$$W_{1OPT} = \sqrt{\frac{I}{P_2}} \times \left\{ \left(B^{1/2}\right)^{-1} \times EigenVector\left( l_{MAX}\left(\left(B^{1/2}\right)^{-1H} A \left(B^{1/2}\right)^{-1}\right)\right)\right\}$$

Appendix 5:

In this appendix, we derive the optimum solution for multiuser case design. It is obvious that the following equation is always true.

$$\sum_{K=1}^{N_{SEC}} \log_2\left(1+I^K \times y^K\right) = \log_2\left(\prod_{K=1}^{N_{SEC}}\left(1+I^K \times y^K\right)\right)$$

Maximizing the left side of the equation is equal to the right side. Furthermore, log-function is a monotonically increasing function of its argument, so maximizing this function is equivalent to maximizing the argument. Therefore, by using the Lagrange method, we derive the optimum solution to the problem below.

$$f(I^1, I^2, \ldots, I^{N_{SEC}}, l) = \prod_{K=1}^{N_{SEC}}\left(1+I^K \times y^K\right) + l\left(I - \sum_{K=1}^{N_{SEC}} I^K\right)$$

By differentiating the above function due to its parameters, the optimum solution would be derived.

$$\begin{cases} \dfrac{\partial f(I^1, I^2, \ldots, I^{N_{SEC}}, l)}{\partial I_K} = 0 \rightarrow y^J \prod_{\substack{K=1 \\ K \neq J}}^{N_{SEC}}\left(1+I^K \times y^K\right) = l \\ \dfrac{\partial f(I^1, I^2, \ldots, I^{N_{SEC}}, l)}{\partial l} = 0 \rightarrow \sum_{K=1}^{N_{SEC}} I^K = I \end{cases}$$

The second equation is obvious. We use the first one to derive the solution:

$$y^{J}\prod_{\substack{K=1\\K\neq J}}^{N_{SEC}}\left(1+I^{K}\times y^{K}\right)=l\rightarrow y^{J}\prod_{\substack{K=1\\K\neq J}}^{N_{SEC}}\left(1+I^{K}\times y^{K}\right)=y^{T}\prod_{\substack{K=1\\K\neq T}}^{N_{SEC}}\left(1+I^{K}\times y^{K}\right)$$

$$\rightarrow\frac{y^{J}}{1+I^{J}\times y^{J}}=\frac{y^{T}}{1+I^{T}\times y^{T}}\rightarrow I^{J}=I^{T}+\left(\frac{1}{y^{T}}-\frac{1}{y^{J}}\right)$$

By inserting the last equation in $\sum_{K=1}^{N_{SEC}} I^{K}=I$, we could derive the solution presented in this paper.

Appendix 6:

Consider the following equation:

$$\log_{2}\left(1+I^{K}\times y^{K}\right)=\log_{2}\left(1+I^{J}\times y^{J}\right)$$

$$\rightarrow I^{K}\times y^{K}=I^{J}\times y^{J}\rightarrow I^{J}=\frac{y^{J}}{y^{K}}I^{K}$$

By inserting the last equation in $\sum_{K=1}^{N_{SEC}} I^{K}=I$, we could derive the solution presented in this paper.

**Reference:**

**[1]** I.F. Akyildiz, W.Y. Lee, M.C. Vuran, S. Mohanty,"**NeXt Generation/Dynamic spectrum Access/Cognitive Radio Wireless Networks: A Survey,''** Computer Networks Journal, (Elsevier), September 2006.

**[2]** FCC Spectrum Policy Task Force, "**Report of the spectrum efficiency working group**," Nov. 2002. Available: http://www.fcc.gov/sptf/files/SEWGFinalReport 1.pdf

**[3]** F. Gao, R. Zhang, Y.-C. Liang, and X. Wang, "**Design of learning based MIMO cognitive radio systems**," IEEE Trans. Veh. Technol., vol. 59, no. 4, pp. 1707-1720, May 2010.

**[4]** J. Mitola and G. Q. Maguire, "**Cognitive radios: making software radios more personal**," IEEE Personal Commun., vol. 6, pp. 13-18, Aug. 1999.

**[5]** S. Haykin, "**Cognitive radio: brain-empowered wireless communications,**" IEEE J. Sel. Areas Commun., vol. 23, no. 2, pp. 201-220, 2005.

**[6]** J. Mitola, "Cognitive radio," Ph.D. thesis, Royal Institute of Technology, (KTH), 2000.

**[7]** Ebrahim A. Gharavol, Ying-Chang Liang, Senior Member, IEEE, and Koen Mouthaan, *Member, IEEE*, "**Robust Downlink Beamforming in Multiuser MISO Cognitive Radio Networks With Imperfect Channel-State Information'"** IEEE TRANSACTIONS ON VEHICULAR TECHNOLOGY, VOL. 59, NO. 6, JULY 2010.

**[8]** S. Yiu, M. Vu, and V. Tarokh, "**Interference reduction by beamforming in cognitive networks**," in *Proc. IEEE GLOBECOM*, New Orleans, LO, Dec. 2008, pp. 1–6.

**[9]** Lan Zhang, Ying-Chang Liang, Yan Xin, and H. Vincent Poor, "**Robust Cognitive Beamforming With Partial Channel State Information** ",IEEE TRANSACTIONS ON WIRELESS COMMUNICATIONS , October 2008.

**[10]** K. Cumanan, R. Krishna, V. Sharma and S. Lambotharan Advanced Signal Processing Group Loughborough University, Leicestershire, UK, LEII 3TU, "**A Robust Beamforming Based Interference Control Technique and its Performance for Cognitive Radios.**", 2008 IEEE.

**[11]** Gan Zheng, *Member, IEEE,* Shaodan Ma, *Member, IEEE,* Kai-Kit Wong, *Senior Member, IEEE,* and Tung-Sang Ng, *Fellow, IEEE,* "**Robust Beamforming in Cognitive Radio**", *IEEE TRANSACTIONS ON WIRELESS COMMUNICATIONS, VOL. 9, NO. 2, FEBRUARY 2010*

**[12]** Ebrahim A. Gharavol, Ying-Chang Liang, Fellow, IEEE, and Koen Mouthaan, Member, IEEE, "**Robust Linear Transceiver Design in MIMO Ad Hoc Cognitive Radio Networks with Imperfect Channel State Information** "IEEE TRANSACTIONS ON WIRELESS COMMUNICATIONS, 2011 IEEE.

**[13]** P. Ubaidolla and A. Chockalingam, "**Robust joint precoder/receiver filter design for multiuser MIMO downlink,**" in Proc. IEEE International Workshop on Signal Processing Advances in Wireless Communications, pp. 136-140, June 2009.

**[14]** N. Vucic, H. Boche, and S. Shi, "**Robust transceiver optimization in downlink multiuser MIMO systems**," IEEE Trans. Signal Process., vol. 57, no. 9, pp. 3576-3587, Sep. 2009.

**[15]** Erik G. Larsson and Petre Stoica, "**Space-Time Block Coding for Wireless Communications**", Cambridge University Press 2003.

**[16]** Mats Bengtsson , Mats Bengtsson. **"Optimum and Suboptimum Transmit Beamforming"** Book Chapter, *Handbook Of Antennas In Wireless Communications*, 2002.

**[17]** Simon Haykin; "Adaptive filters" 2001

**[18]** Online available at {www.cvxr.com}